\documentclass[11pt,a4paper]{article}
\usepackage{jcappub}
\usepackage{graphicx}
\usepackage{dcolumn}
\usepackage{bm}
\usepackage{natbib}
\usepackage{color}
\fontencoding{T1}

\def\araa{ARA\&A}%
\def\apj{ApJ}

\title{On the Validity of Cosmological Fisher Matrix Forecasts}
\author[1,2,3]{Laura Wolz,}
\author[4,2,3]{Martin Kilbinger,}
\author[2,3,5]{Jochen Weller}
\author[2,3]{and Tommaso Giannantonio}

\affiliation[1]{Department of Physics and Astronomy, University College London, London WC1E 6BT, UK}
\affiliation[2]{Universit\"ats-Sternwarte, Ludwig-Maximilians-Universit\"at M\"unchen, Scheinerstr.~1, 81679 M\"unchen, Germany}
\affiliation[3]{Excellence Cluster Universe, Boltzmannstr.~2, 85748 Garching, Germany}
\affiliation[4]{CEA Saclay, Service d'Astrophysique (SAp), Orme des Merisiers, B\^at.~709, 91191 Gif-sur-Yvette, France}
\affiliation[5]{Max-Planck-Institut f\"{u}r extraterrestrische Physik, Giessenbachstr., 85748 Garching, Germany}

\emailAdd{lwolz@star.ucl.ac.uk}
\emailAdd{martin.kilbinger@cea.fr}
\emailAdd{jochen.weller@usm.lmu.de}
\emailAdd{tommaso@usm.lmu.de}

\abstract{
We present a comparison of Fisher matrix forecasts
  for cosmological probes with Monte Carlo Markov Chain (MCMC)
  posterior likelihood estimation methods. We analyse the performance
  of future Dark Energy Task Force (DETF) stage-III and stage-IV
  dark-energy
  surveys using supernovae, baryon acoustic oscillations
  and weak lensing as probes. We concentrate in particular on the
  dark-energy equation of state parameters $w_0$ and $w_a$.
 For purely geometrical probes, and especially when marginalising over
  $w_a$, we find considerable disagreement between the two methods, since in this case the
  Fisher matrix can not reproduce the highly non-elliptical shape of
  the likelihood function. More quantitatively, the Fisher method
  underestimates the marginalized errors for purely geometrical probes
  between 30$\%$-70$\%$. For cases including structure formation such
  as weak lensing, we find that the posterior probability contours
  from the Fisher matrix estimation are in good agreement with the
  MCMC contours and the forecasted errors only changing on the $5\%$ level.
We then explore non-linear transformations
  resulting in physically-motivated parameters and investigate whether
  these parameterisations exhibit a Gaussian behaviour.
 We conclude
  that for the
  purely geometrical probes and, more generally, in cases
    where it is not known whether the likelihood is close to Gaussian, the Fisher matrix is not
  the appropriate tool to produce reliable forecasts.
}

\keywords{Dark
  energy experiments, Supernovae Type Ia, Baryon acoustic
  oscillations, Gravitational lensing, Cosmological parameters from
  LSS}

\notoc

\begin{document}
\maketitle
\flushbottom

\section{Introduction}

Forecasting cosmological observations has in recent
years become an important contribution to the design of future surveys, in
particular with respect to revealing the nature of the observed cosmic
acceleration. Within this context, the Dark Energy Task Force \cite[hereafter
DETF]{DETF:06} compared various stages of present and future surveys and cosmological
probes. They proposed a figure of merit (FoM), which allows one to compare these surveys
in their ability to constrain the equation of state of dark energy in
the so-called Chevallier-Polarski-Linder (CPL) parameterisation \cite{Chevallier:01,Linder:03}. 

In most
cases, such forecast exercises exploit a second-order approximation to
the likelihood, where the parameter covariance is given by the Fisher
information matrix \cite{KS69, TTH97, Tegmark:1997rp}. This is usually the method of
choice, because it is fast and easy to implement, and the results of
which are straightforwardly reproduced. However, since the Fisher
matrix is a local approximation of the likelihood, its application
can lead to a spurious breaking of some --- even infinite ---
degeneracies. This has been long known, for example in the case of
cosmic microwave background (CMB) anisotropies. If one neglects the
integrated Sachs-Wolfe (ISW) effect on large scales, the CMB can not
simultaneously constrain the dark-energy parameters and the curvature of
the Universe. However, a Fisher matrix approach would give relatively
tight constraints on quantities which in reality are infinitely
degenerate.

 The DETF provides a
prescription to circumvent this problem: instead of using the
parameters of dark energy and curvature directly, they propose to take as a
variable the angular size of the sound horizon at last scattering, which is well constrained, and whose associated likelihood
distribution is approximately Gaussian. The parameter space is then
expanded with a suitably chosen Jacobian to its full
dimensionality. This by construction leads to the desired, infinite
degeneracies \cite{DETF:06,Rassat:2008p2048}. Interestingly, the
angular size of the sound horizon is also a `good' variable to explore
the posterior likelihood efficiently with Monte Carlo methods
\cite{2002PhRvD..66f3007K,2002PhRvD..66j3511L}.
 
Prominent cosmological probes other than the CMB are the large-scale
distribution of galaxies and the baryon acoustic oscillation (BAO) feature
in the power spectrum \cite{2003ApJ...598..720S}, weak cosmological
lensing (WL) or cosmic shear \cite{2008ARNPS..58...99H,
  2010CQGra..27w3001B}, the spatial and redshift distribution of
galaxy clusters \cite{2011ARAandA..49..409A} and the
magnitude-redshift relation as measured with Type Ia Supernovae (SN)
\cite{Perlmutter:97,Riess:98,2008A&A...486..375A}. All these probes measure cosmological
parameters with either geometric distance measures, the growth of
structures, or a mixture thereof \cite{2012arXiv1201.2434W}.  However,
for these probes the Gaussian nature of the likelihood functions is
less explored than in the case of CMB anisotropies. Recently,
Ref.~\cite{Joachimi:2011iq} presented a study in the context of cosmic
shear forecasts, which allows to treat mildly non-Gaussian
distributions by non-linear parameter transformations.

A more robust approach to the forecast problem is possible, by sampling the full likelihood function without approximations, as was recently used e.g. by Refs.~\cite{Perotto:2006,Martinelli:2010wn,DeBernardis:2011iw}.
In order to assess the question of how valid a standard Fisher matrix
approach is in forecasting constraints, we investigate a selection of
future probes, comparing for each case a direct likelihood
exploration using a Monte Carlo Markov Chain (MCMC) approach with the
Fisher matrix approximation. We will also describe possible alternative
parameterisations in which the likelihood takes an approximate
Gaussian shape and hence becomes more suitable to a Fisher matrix
approach.
 
This study will be of importance for forecasts for future probes. In
the coming decade, many new and ever larger surveys are planned and
carried out, like
Pan-STARRS\footnote{\texttt{http://pan-starrs.ifa.hawaii.edu/public}},
KiDS\footnote{\texttt{http://www.astro-wise.org/projects/KIDS}},
DES\footnote{\texttt{www.darkenergysurvey.org}},
HETDEX\footnote{\texttt{http://hetdex.org}},
Euclid\footnote{\texttt{www.euclid-ec.org}},
WFIRST\footnote{\texttt{http://wfirst.gsfc.nasa.gov}},
LSST\footnote{\texttt{http://www.lsst.org/lsst}} and
SKA\footnote{\texttt{http://www.skatelescope.org}}.

The paper is organised as follows. In Section \ref{sec:likelihood} we
briefly describe the methods to explore the likelihood with a MCMC
method and the Fisher matrix approximation. Section
\ref{sec:cosmology} presents the cosmological probes and surveys we
investigate. In Section \ref{sec:gauss} we sketch a possible
alternative parameterisation to obtain a more Gaussian likelihood, followed by
a discussion and our conclusions in Section~\ref{sec:conclusion}.

\section{Forecasting: Fisher matrix vs. Monte Carlo sampling}\label{sec:likelihood}

To investigate the distribution of a set of
 parameters
$\boldsymbol\theta$ under the assumption of a certain (cosmological) model $M$ and given the data
$\mathbf D$, we have to estimate
the posterior probability $\mathcal{P(\boldsymbol \theta)} \equiv p(\boldsymbol \theta | \mathbf D, M)$. This is given by Bayes' theorem as a function of the likelihood  $\mathcal{L(\boldsymbol \theta)} \equiv p( \mathbf D | \boldsymbol \theta, M)$, the prior $\Pi (\boldsymbol \theta) \equiv p(\boldsymbol \theta | M) $ and the Bayesian evidence $\mathcal{Z} \equiv p( \mathbf D | M)$, as
\begin{equation}
\mathcal{P(\boldsymbol \theta)} = \frac{\mathcal{L(\boldsymbol \theta)} \, \Pi (\boldsymbol \theta)} {\mathcal{Z}} \, .
\end{equation}
If we are interested in parameter estimation for a fixed model only, then the evidence $\mathcal Z$ is a trivial normalising factor and can be neglected. The priors are likewise irrelevant for our discussion, and we will take them to be flat in the chosen parameterisation. 
The likelihood function can be sampled in principle by scanning the entire parameter space on a grid and calculating
its corresponding values. However, this is very inefficient and
for high-dimensional parameter spaces not feasible. An efficient
method to circumvent this problem is the Monte Carlo sampling of
the likelihood with the help of a Markov chain.

\subsection{MCMC implementation}

We implement a Monte Carlo Markov Chain by using the
Metropolis-Hastings algorithm \cite{Metropolis53, Hastings70}.  Using a Monte-Carlo integration,
the expectation of any function $f(\boldsymbol\theta)$ can be estimated as
\begin{equation}
  \int f(\boldsymbol\theta) \, \mathcal{P(\boldsymbol \theta)} \, {\rm d}^n\theta \simeq \frac 1 N
  \sum_{i=1}^N f(\boldsymbol\theta^{(i)}) \label{montecarloapprox},
\end{equation}
where the Markov Chain points $\{\boldsymbol \theta^{(i)}\}_{i=1}^N$ represent a sample under
the posterior $\mathcal{P}(\boldsymbol \theta)$.

Given the point in the chain $\boldsymbol\theta^{(i)}$, a new point $\boldsymbol\theta^\star$ is drawn from a distribution $\psi(\boldsymbol\theta^{(i)} - \boldsymbol\theta^\star)$. The new point is accepted as $\boldsymbol\theta^{(i+1)}$ with the probability
min$\{1, \mathcal{P}(\boldsymbol\theta^\star) / \mathcal{P}(\boldsymbol\theta^{(i)})  \}$.
We choose as the so-called proposal density $\psi$
a multi-variate Gaussian with a
diagonal covariance.
As a criterion for the convergence of the chains we implement the Gelman-Rubin test
\cite{Gelman:92} by running five independent chains with
over-dispersed seeds. Calculating $R$, which is proportional to the ratio of the parameter variances of one chain to the mean parameter variances of the independent chains, yields
a quick check of convergence. We consider a chain to be converged when
$\sqrt{R}-1<0.05$.

We run MCM chains for each of the cosmological probes: supernovae, BAO and weak lensing, as described in Section~\ref{sec:cosmology}. The chains are then analysed by investigating the
two-dimensional joint marginalised posterior distributions, and compared to the Fisher matrix results.

\subsection{Fisher matrix formalism}

For a posterior distribution $\mathcal{P(\boldsymbol \theta)} $ the
Fisher information matrix
\cite{KS69, TTH97} is defined as the expectation value
\begin{equation}
  F_{\alpha \beta} = \left\langle \frac{\partial^2
    \left[ - \ln \mathcal{P}(\boldsymbol \theta)  \right]}{\partial \theta_\alpha \, \partial \theta_\beta} \right\rangle.
\end{equation}
We use wide uniform priors on the parameters throughout, therefore
the notions of posterior and likelihood are interchangeable.

The inverse of the Fisher matrix is the curvature of the likelihood evaluated at the
mean.
Usually, the mean is assumed to be equivalent to the maximum likelihood (ML) estimator. $F^{-1}$ is
thus a local measure of how fast the likelihood falls off from the
maximum in different directions. According to the Cram\'er-Rao
inequality, the Fisher matrix gives an upper bound on the parameter
error $\Delta$ on a parameter $\theta_\alpha$,
\begin{equation}
  \Delta \theta_\alpha \leq \sqrt{\left( F^{-1} \right)_{\alpha \alpha}}.
  \label{cramer_rao}
\end{equation}
These two properties make the Fisher matrix a first-order, optimistic
approximation of the likelihood. It is calculated very quickly, in
particular compared to estimating or sampling the full posterior
distribution, if the dimensionality of the parameter space is
high.
Therefore, the Fisher matrix is used in most cases of parameter
forecasts, such as e.g. for predictions of the performance of planned surveys, which play a
crucial role in defining the science case and informing funding
agencies.

By construction, the Fisher matrix is expected to give, more or less, inaccurate results for
non-Gaussian posteriors. In this case, the predicted errors will most
likely be too small, and parameter degeneracies could be
misestimated, since they are only considered to be linear. This
might lead to regions in parameter space to be spuriously excluded,
where in fact they are permitted by the full posterior, and vice-versa.

\section{Cosmological Probes}\label{sec:cosmology}

Some of the major cosmological probes which in recent years have
tested scenarios of cosmic acceleration are the observation of the
cosmological shearing of galaxy shapes \cite{SHJKS09}, the large-scale
distribution of galaxies and the associated baryon acoustic feature
\cite{2011MNRAS.416.3017B, 2010MNRAS.404...60R, 2012arXiv1201.2137H},
and the magnitude-redshift diagram of SN of type Ia
\cite{2008A&A...486..375A, 2009ApJS..185...32K, 2011ApJ...737..102S, 2012ApJ...746...85S}.
In this work, we model forecasts of these three probes for future
surveys.  The following section briefly describes the surveys. Details
of the settings can be found in Table~\ref{tab:surveysettings}.

Our cosmological model is a flat $w$CDM model with
a dark-energy equation of state evolving linearly with the scale
factor $a$, $w(a) = w_0 + w_a (1 - a)$ \cite{Chevallier:01,
  Linder:03}. We vary the total matter density $\Omega_{\rm m}$ and the two
dark-energy parameters, using the fiducial model $\{ \Omega_{\rm m}, w_0, w_a\} = \{0.3, \, -1, \, 
0\}$.
For weak lensing, we include $\sigma_8$ in the analysis, with fiducial
value $\sigma_8 = 0.8$. We keep fixed the baryon density $\Omega_{\rm b}
= 0.0462$, the Hubble parameter $h = 0.72$ and the spectral index of
primordial density perturbations $n_{\rm s} = 0.96$, corresponding to the WMAP7 values \cite{2010arXiv1001.4538K}.

\subsection{The surveys}

The Dark Energy Survey (DES) \cite{2005astro.ph.10346T} will
observe 5,000 square degrees in the Southern sky, starting in late
2012.  This DETF stage-III imaging survey consists of the five optical
bands $g,r,i,z$ and $Y$, using the newly-built DECam on the 4m Blanco
telescope at the Cerro Tololo Inter-American Observatory (CTIO). The
four main cosmological probes of DES are weak gravitational lensing,
baryon acoustic oscillations, clusters of galaxies and Type Ia supernovae.

The Hobby-Eberly Telescope Dark Energy Experiment (HETDEX)
\cite{Hill:2008mv} is a deep optical spectroscopic survey
designed to measure the expansion rate, the angular diameter distance
and the growth of structure in the redshift range $2 < z < 4 $ to high
accuracy. It will observe 420 square degrees on the Northern hemisphere
and measure the redshift of 0.8 million Ly$_\alpha$-emitting
galaxies. HETDEX is a DETF stage-III survey, expected to
be completed by the end of 2014.

Euclid \cite{2011arXiv1110.3193L} is a satellite with a 1.2m mirror,
scheduled as medium-class mission by the European Space Agency (ESA)
for launch in 2019. This stage-IV survey will observe 15,000 square
degrees of extra-galactic sky with optical and near-infrared imaging,
and near-infrared spectroscopy. We will consider here the photometric part only; the imaging filters consist of one
broad optical band ($R+I+Z$) and the three infrared filters $Y, J$ and
$H$. To obtain photometric redshifts, Euclid will be complemented by
ground-based optical surveys such as DES or Pan-STARRS.  The
two main science drivers for Euclid are weak lensing and galaxy
clustering (BAO and redshift-space distortions) as cosmological probes.

The planned radio interferometer Square Kilometre Array
(SKA) 
\cite{Blake:2004pb,Rawlings:2011dd} is a stage-IV survey designed to
measure the 21-centimeter spectral line of neutral hydrogen (HI). SKA's key features
are to access the epoch of reionisation and to probe the large-scale structure
of HI-emitting galaxies on a sky area of 20,000 square degrees. 

\begin{table*} 
\begin {center}

\begin{tabular}{|c|c|c|c|c|c|c|c|c|}
\hline
Stage & Experiment & $\Omega_{\rm sky} \, [\mathrm{deg}^2]$ & $z_{\min}$ &
$z_{\max}$ & $z$ bins & $ {\langle |\gamma|^2 \rangle}^{1/2}$ & gal. dens. & $z_{\rm med}$\\
\hline
III & DES    & 5,000  &  0.01 & 2.0 &  5 & 0.16  & 12 $\mathrm{arcmin}^{-2}$ & 0.8 \\
III & HETDEX & 420   &  1.8  & 3.9 & 10 &  ---    & ---    & --- \\
IV & Euclid (photo) & 15,000 &  0.01 & 2.5 &  8 & 0.247 & 30 $\mathrm{arcmin}^{-2}$ & 0.8 \\
IV & SKA    & 20,000 &  0.01 & 1.5 & 10 &  ---    & ---    & --- \\
\hline
\end{tabular}
\caption{Survey settings used for producing the forecasts.}
\label{tab:surveysettings}
\end{center}
\end{table*}

\subsection{Supernovae}

Supernovae of type Ia are explosions of white dwarf stars at or near
the Chandrasekhar mass limit of  $\sim 1.4 \, M_{\odot}$. The uniform
progenitor mass makes them standardisable candles, and they can
therefore be used as geometrical measures for cosmological models after rescaling for the stretch of their light curves \cite{Phillips:1993a}. In this work, a
supernova survey with 1,000 objects equally spaced in the redshift
range $0 < z < 1.8$ is taken as reference; this scenario corresponds
to a DETF stage-IV forecast. The observed magnitudes are modeled, ignoring for our purposes the effects of reddening and higher-order stretch corrections, as \cite{Perlmutter:97}
\begin{equation}
  m(z) = M_{\rm int} + 5 \, \log_{10} \left[ H_0\, d_{\rm L}(z) \right] \, ;
\end{equation}
here $M_{\rm int}$ is the intrinsic magnitude of the supernovae and the luminosity distance is
\begin{equation}
d_L(z) = c \, (1+z) \, \int_0^z \frac{dz'}{H(z')} \;,
\end{equation}
where the Hubble expansion rate is given for our parameterised dark-energy model by:
\begin{equation}
 \frac{H(z)}{H_0} =  \sqrt{\Omega_{\rm m} (1+z)^{3} +  \Omega_{\Lambda} e^{-3w_a\frac z{1+z}}(1+z)^{3(1+w_0+w_a) }}\;.
 \label{Eq:expansionrate}
\end{equation}
Note that we fix $H_0$ for the likelihood analysis because it is completely degenerate with the intrinsic magnitude.

\begin{figure*}
\begin {center}
\includegraphics[width=0.32\textwidth, clip, trim=0mm 0mm 40mm 0mm]{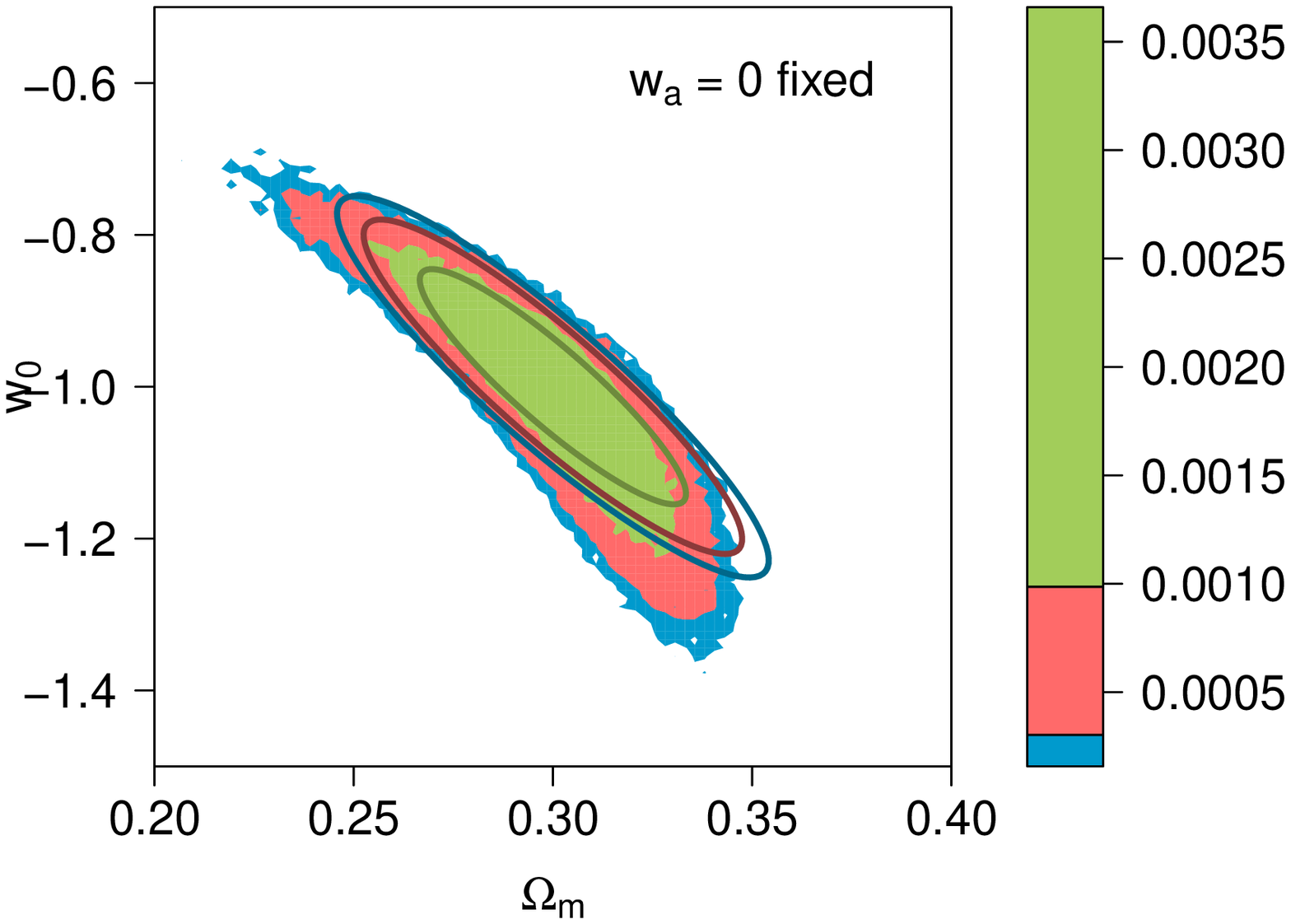}%
\includegraphics[width=0.32\textwidth, clip, trim=0mm 0mm 40mm 0mm]{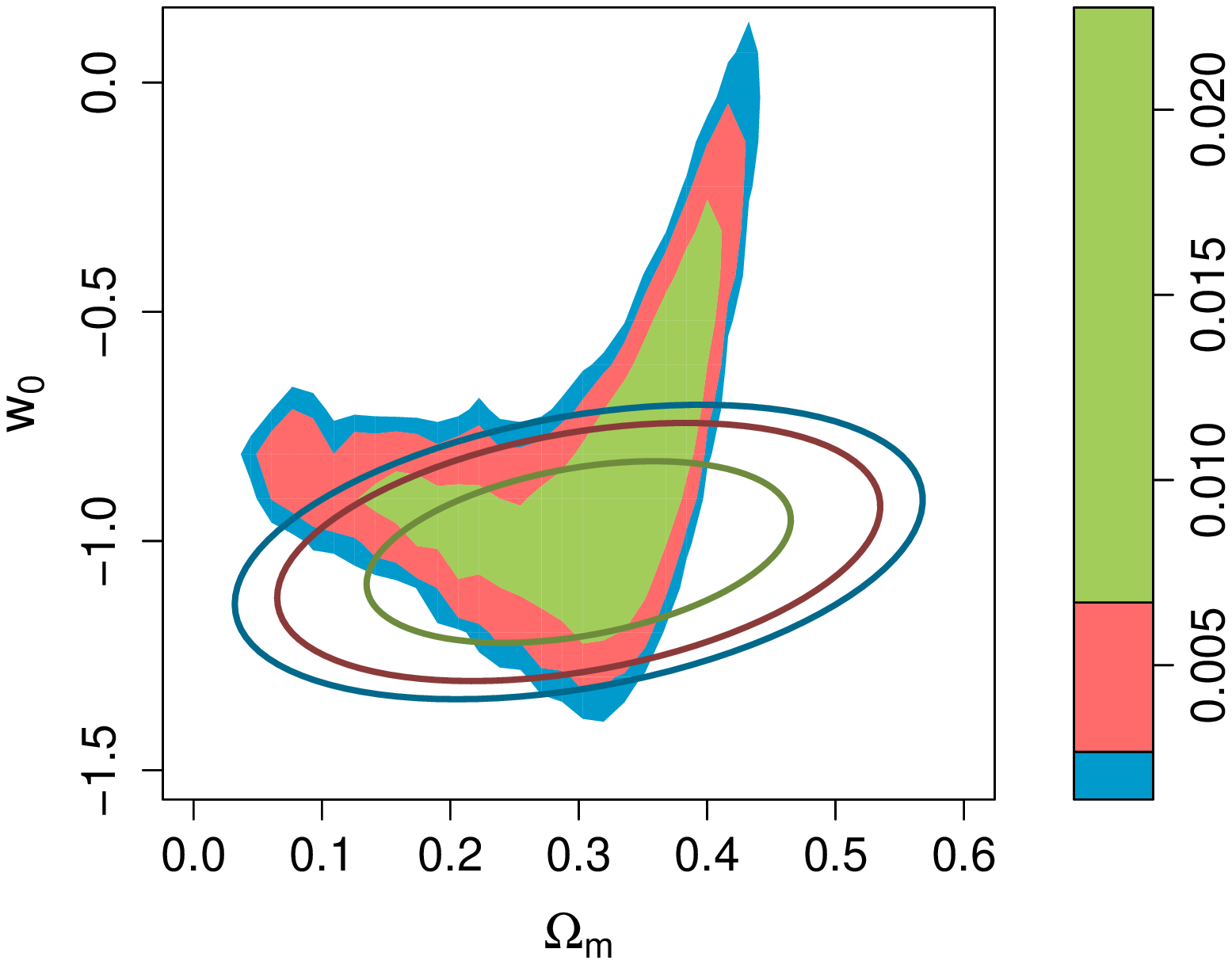}
\includegraphics[width=0.32\textwidth, clip, trim=0mm 0mm 40mm 0mm]{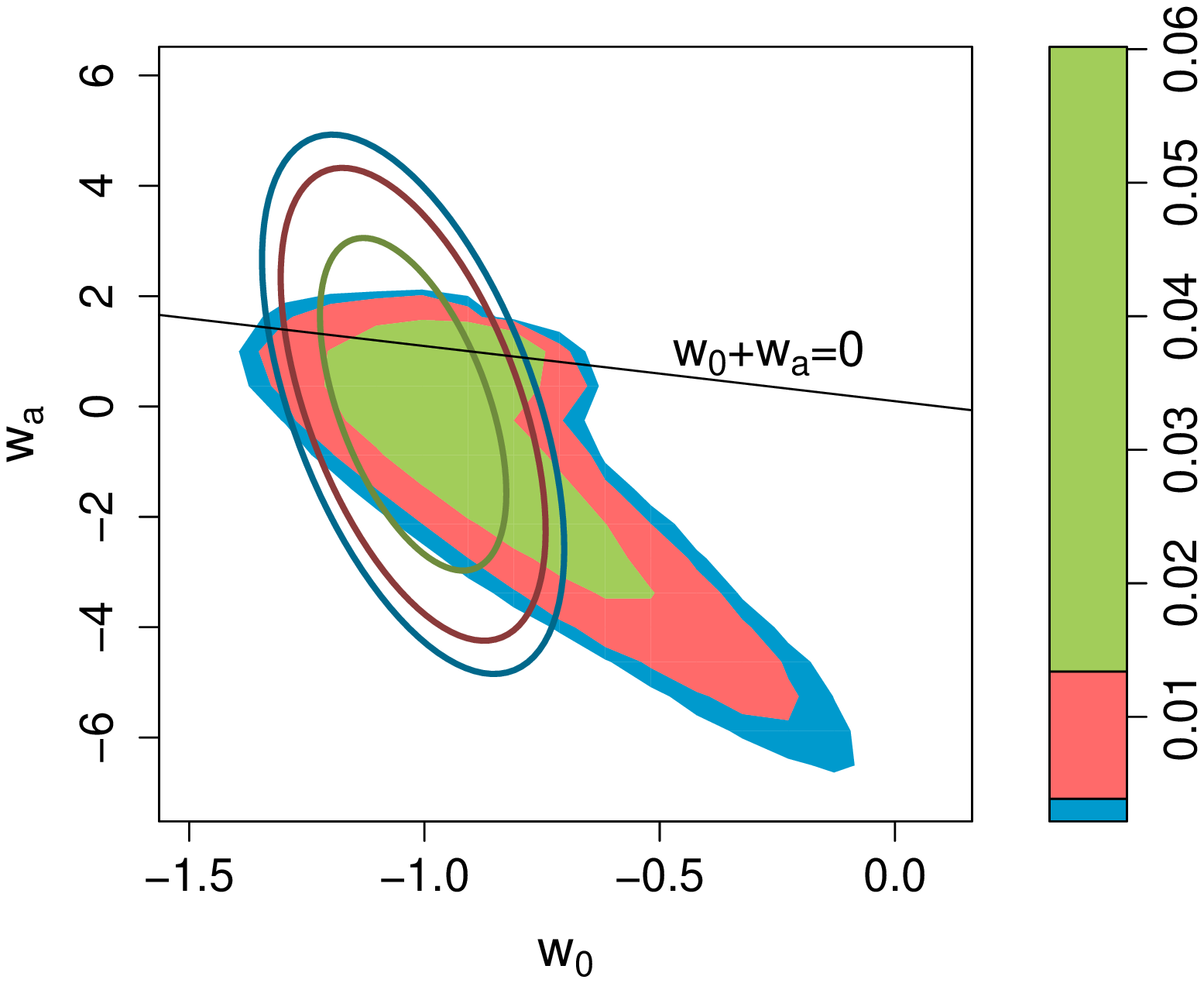}
 \caption{68\%, 90\% and 95\% confidence regions for a supernova
   survey. Filled contours correspond to the full posterior sampled
   with MCMC, while the solid lines represent the Fisher matrix
   results. The parameter spaces are $\{\Omega_{\rm m}, w_0, M_{\rm int}\}$
   with fixed $w_a = 0$
   (\emph{left panel}), and $\{\Omega_{\rm m}, w_0, w_a, M_{\rm
     int}\}$ (\emph{middle and right panels}).
   The parameters which are not shown have been marginalised in all panels.
 }

 \label{SN_omm_w0_notmarg}
\end{center}
\end{figure*}

To calculate the likelihood, we independently draw 1,000 supernova
magnitudes from a Gaussian distribution with variance $\sigma_{m} = 0.15$.
We consider the case of a fixed $w_a = 0$ for the likelihood analysis,
and the case where $w_a$ is fit jointly with the other cosmological
parameters. Here we marginalise over the modified intrinsic magnitude ${\cal M}=M_{\rm int}-5\log H_0$.
The results are displayed in
Fig.~\ref{SN_omm_w0_notmarg}, showing two-dimensional marginals from MCMC compared to the Fisher matrix ellipses. In the left panel, where we fix $w_a = \mathrm{const} = 0$, the Fisher matrix result for the
$\Omega_{\rm m}$--$w_0$ joint likelihood is a fair approximation of the full likelihood. However, it is substantially smaller, in particular at the $ > 90\%$ level. 
This is in strong contrast to the second case shown in the middle panel:
when marginalising over $w_a$, the Fisher prediction is completely
misleading. The curvature of the likelihood function is highly
variable over the parameter space, and the corresponding bends in the contours cause
very different parameter regions to be covered compared to the Fisher
ellipsoid. We will explain the parameter degeneracies in more detail
in Section~\ref{sec:BAO}, describing the similar behaviour for the BAO forecasts.  
The right panel of Fig.~\ref{SN_omm_w0_notmarg} shows the comparison of the techniques in the $w_0$--$w_a$
space. Although the total sizes of the contours do not differ
significantly, the Fisher matrix fails to predict the physical boundary
$w_a+w_0<0$ which is required to explain the accelerated expansion in
the dark-energy cosmological model. Therefore, the full posterior
shifts towards the negative range in $w_a$. %

\subsection{Baryon acoustic oscillations}
\label{sec:BAO}

The primordial density fluctuations in the early Universe leave an
imprint on the dark matter and galaxy distributions, which can be still
observed at much later times. Baryon acoustic oscillations (BAOs)
cause an enhancement of the galaxy correlation function at a comoving distance of about 150
Mpc. This feature can be used as standard ruler, and
therefore serves as a measure of the geometry of the Universe \cite{Eisenstein:05,2005MNRAS.362..505C,2012arXiv1203.6594A}.

In an ideal case, to encompass all the information of a BAO survey, the full
three-dimensional power spectrum or correlation function should be used for constraining
cosmology. The analysis of the long-wavelength power spectrum requires
careful modeling, and can increase the constraining power on the cosmological parameters
\cite{2008MNRAS.390.1470S,2012MNRAS.tmp.2584M}. Here, we choose a simplified ansatz
and model a spectroscopic BAO survey following Refs.~\cite{Blake:2006nx} and
\cite{DETF:06}: this assumes that the observations yield information on the angular
and line-of-sight comoving distances. Accordingly, the observables to enter the
likelihood are the logarithms of the comoving angular diameter distance $d_{\rm A}(z)
= d_{\rm L}(z) / (1+z)^{2}$ and the Hubble parameter $H(z)$. The statistical error of
the former is
\begin{equation}
\sigma_{d_{\rm A}}(z) = \begin{cases}
0.011 \sqrt{\frac{V_0}{V(z)}} \left(\frac{z_{\rm m}}{z}\right)^{0.5} &z<z_{\rm m}\\
0.011 \sqrt{\frac{V_0}{V(z)}} & z>z_{\rm m}, 
\end{cases}
\end{equation}
where the different cases refer to the non-linear evolution of matter for $z<z_{\rm
  m} = 1.4$. $V(z) = d^2_{\rm A}(z) H^{-1}(z) \Omega_{\rm sky} \Delta z$ is
the comoving volume for a redshift bin with width $\Delta z$, and
observed solid angle $\Omega_{\rm sky}$. The reference volume is $V_0
= 2.16 \, h^{-3}$ Gpc. The statistical error of the Hubble parameter is modeled
as $\sigma_{H} = 1.74 \, \sigma_{d_{\rm A}}$. We add in
quadrature a systematic contribution of $\sigma_{\rm sys} = 0.01
\sqrt{0.5 / \Delta z}$ to both errors. Typically, BAO forecasts use
some combination of $H(z)$ and $d_A(z)$ as parameters, and then
linearly transform to the dark-energy parameters $w_0$ and $w_a$
\cite{Seo:05}. Strictly speaking, it is the validity of this linear
transformation we investigate here:
 by construction the likelihoods on
 $H(z)$ and $d_A(z)$ are Gaussian. Since we use a Fisher matrix
 formalism, the errors on $w_0$ and $w_a$ follow from a linear
 transformation and are therefore also Gaussian. However,
 the relation between $H(z)$ and $d_A(z)$ is truly non-linear and the
 Fisher matrix formulation can not encompass this behaviour.

\begin{figure*}
\begin {center}

  \includegraphics[width=0.32\textwidth, clip, trim=0mm 0mm 40mm 0mm]{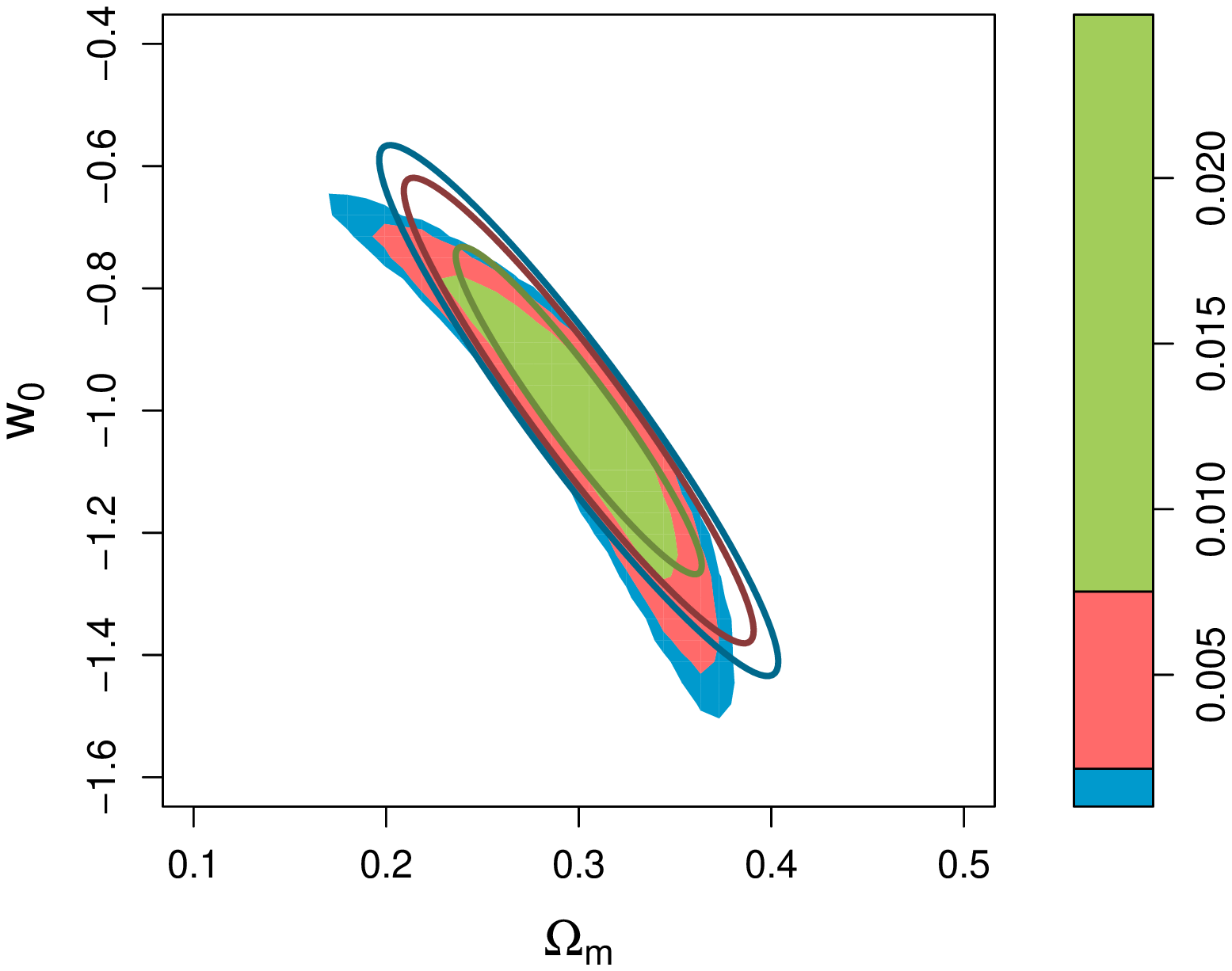}%
  \includegraphics[width=0.32\textwidth, clip, trim=0mm 0mm 40mm 0mm]{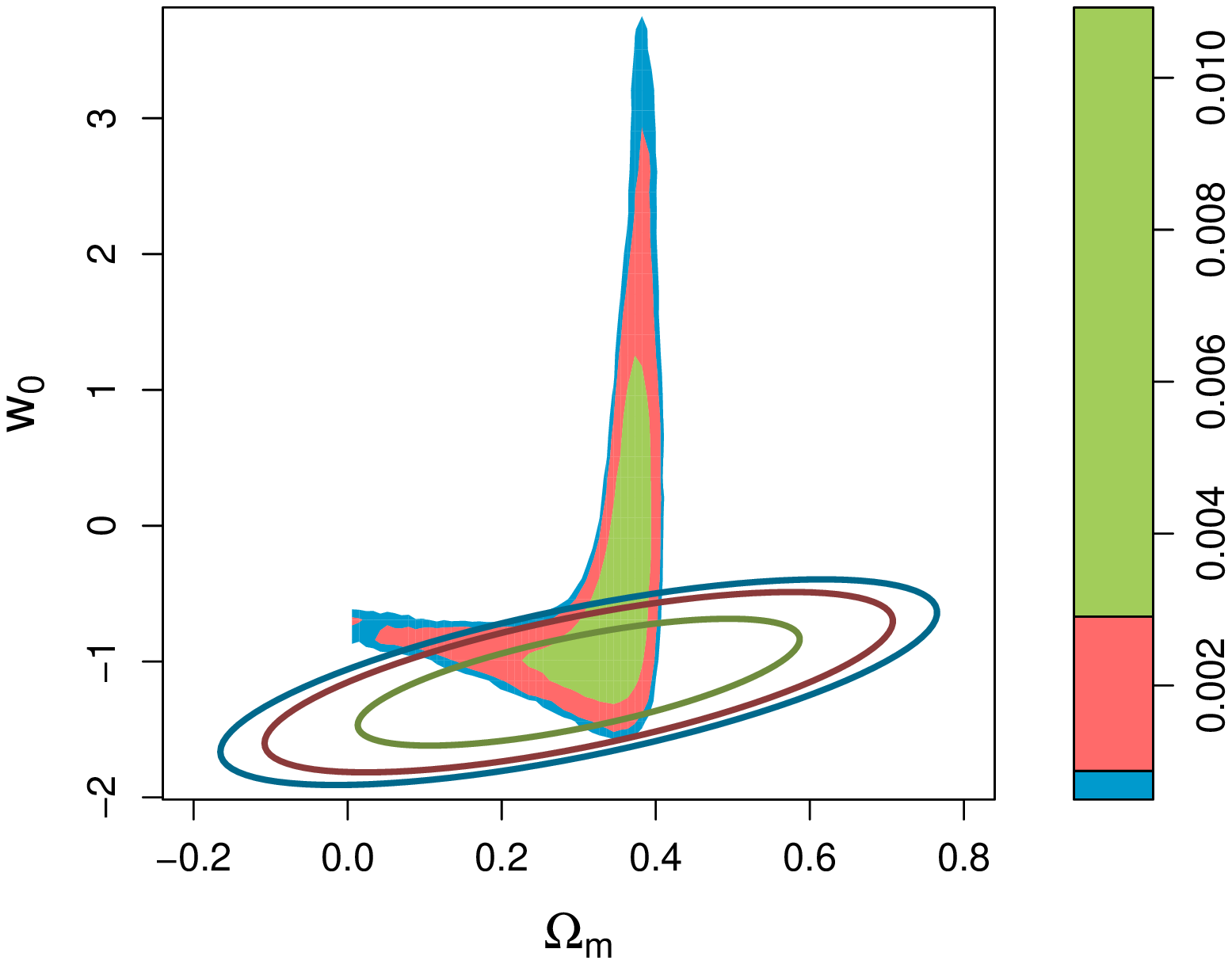}%
 \includegraphics[width=0.32\textwidth, clip, trim=0mm 0mm 40mm 0mm]{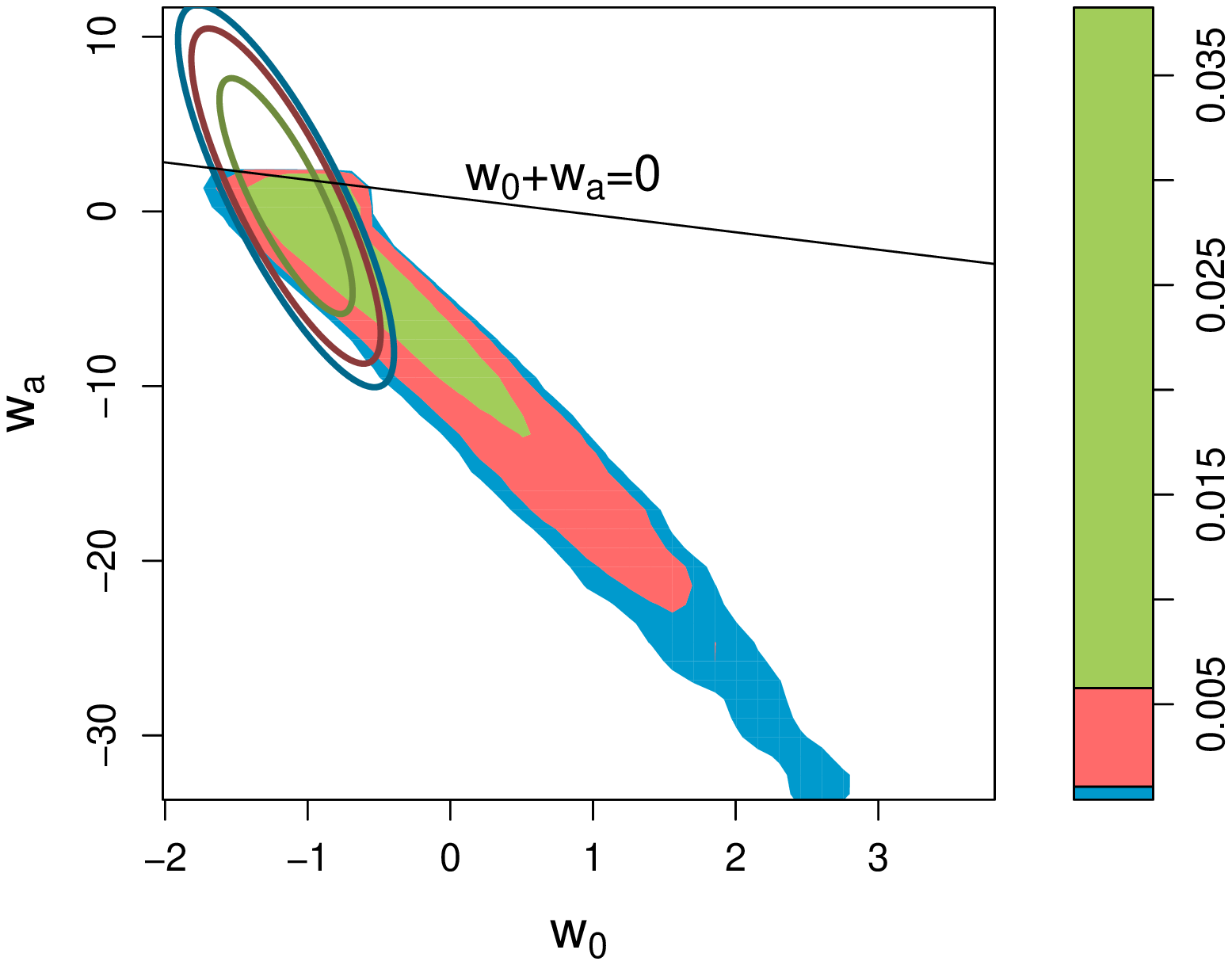}%
\hfill%

  \caption{Confidence regions for a SKA-like BAO survey. Filled contours correspond to the MCMC, while the solid lines represent the Fisher matrix
   results. The
   parameter space is $\{\Omega_{\rm m}, w_0\}$ (\emph{left panel}),
   and $\{\Omega_{\rm m}, w_0, w_a\}$ (\emph{middle and right
     panels}). In the latter cases, we marginalise over the hidden parameter.
}

  \label{BAO_SKA_w0_wa}
\end{center}
\end{figure*}

\begin{figure}
\begin{center}
\includegraphics[width=0.49\textwidth]{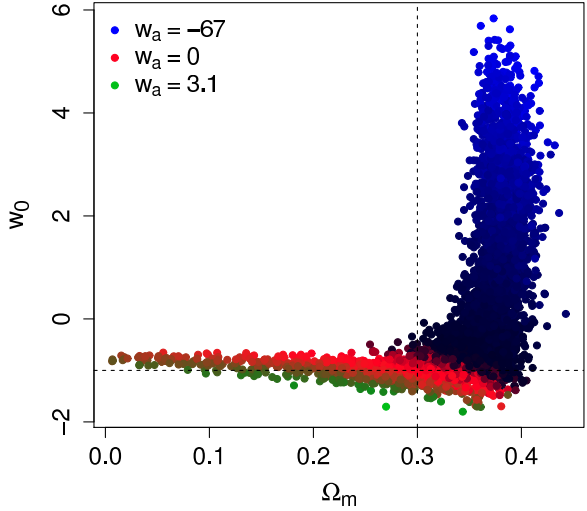}
\caption{Plot of the accepted Markov chain sample points for a SKA-like BAO
  survey. The colour indicates the value of $w_a$, going from $-67$
  (dark blue) to $3.1$ (green). }
\label{BAO_SKA_3d}
\end{center}
\end{figure}

\begin{figure*}
\begin{center}
\includegraphics[width=0.45\textwidth]{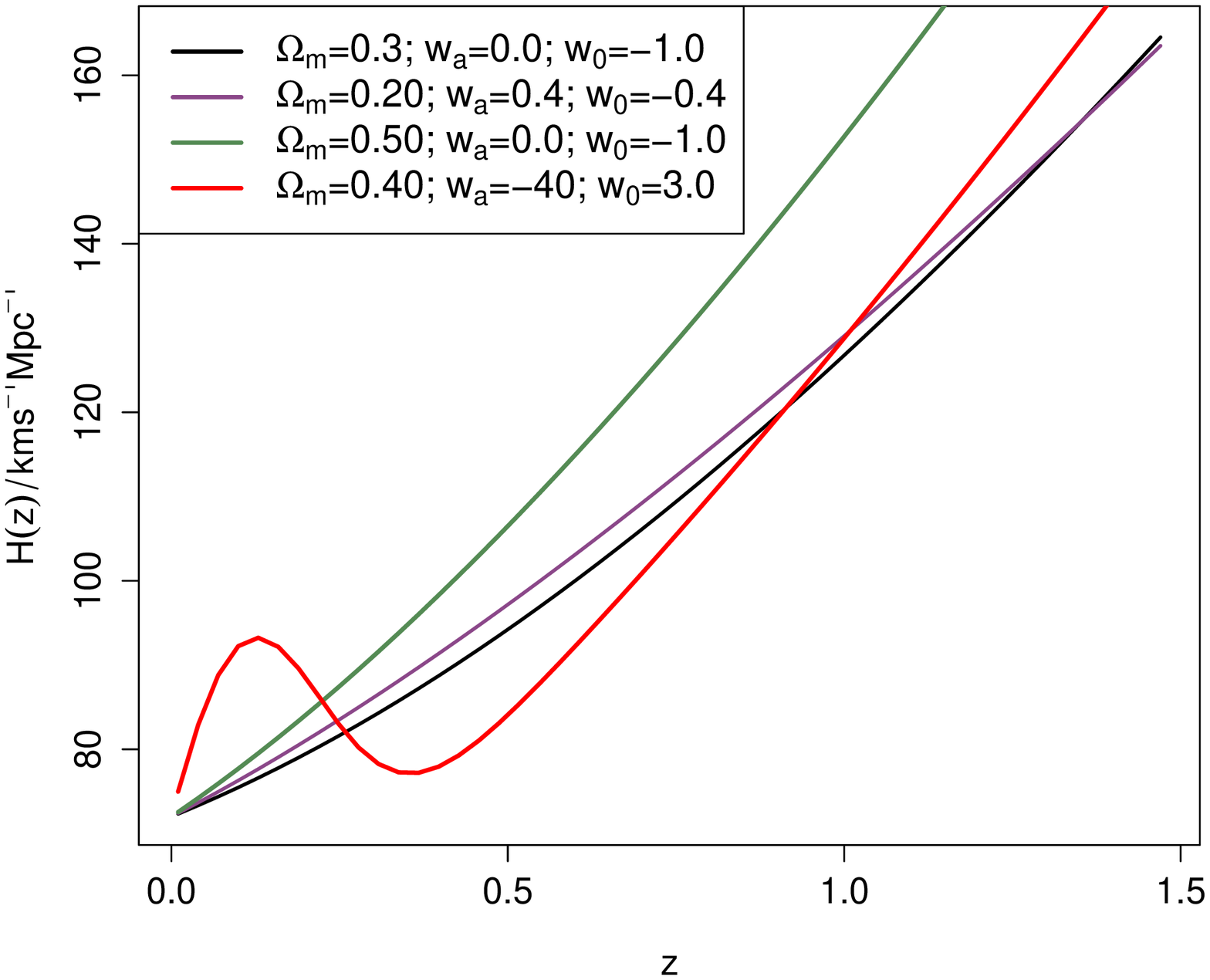}
\includegraphics[width=0.45\textwidth]{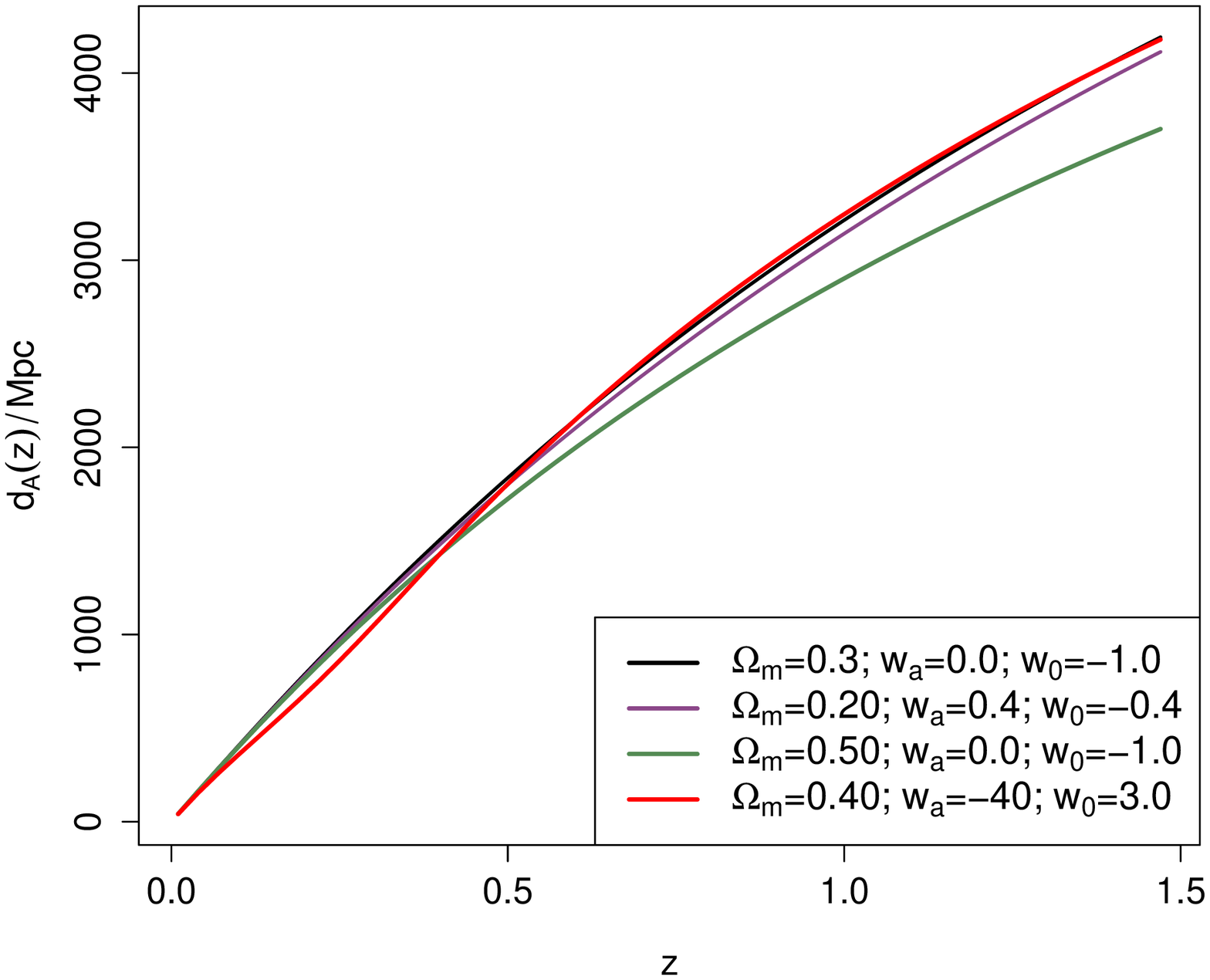}
\caption{Expansion rate $H(z)$ (\emph{left}) and comoving angular
  diameter distance $d_{\rm A}(z)$ (\emph{right}) versus redshift for different cosmological models.}
\label{BAO_models_comp}
\end{center}
\end{figure*}

As for supernovae, we consider the two cases of $w_a$ being a fixed
and a free
parameter in the analysis, respectively. The results for a stage-IV,
SKA-like survey, shown in Fig.~\ref{BAO_SKA_w0_wa}, are similar to the SNIa case: for a constant $w_a$, the likelihood
function is only slightly bent, and the Fisher matrix is a reasonable
approximation. When including a free $w_a$ however, the likelihood function becomes
highly non-Gaussian, and its shape cannot be described any more by the curvature at
the maximum only. 
This is an example where the variance and figure of merit from
the Fisher matrix would suggest larger constraints than using the full
likelihood. The reason for this is the relatively broad, low-curvature region
around the maximum likelihood, which is captured by the Fisher matrix; the true likelihood
function however is falling off quickly and bending away from the initial direction,
excluding large regions covered by the Fisher ellipse. This behaviour becomes clearer when considering Figs.~\ref{BAO_SKA_3d} and \ref{BAO_models_comp}.
The sample cloud pictured in Fig.~\ref{BAO_SKA_3d} shows the
three-dimensional distribution of the MCMC. The
$w_a$ value of the sample points is colour-coded, transitioning from green to blue with
$w_a$ going from positive to negative. The plot shows that there is a
non-negligible correlation between the dark-energy parameters, namely
positive $w_0$ and highly negative $w_a$ values, which is consistent with the condition $w_0+w_a<0$. 
Fig.~\ref{BAO_models_comp} illustrates that some cosmological models with a highly non-standard expansion rate (red curve) generate angular diameter distances which are similar to the reference one with our LCDM fiducial parameters (black curve). 
This degeneracy of the CPL dark-energy parameters for geometrical probes originates from Eq.~(\ref{Eq:expansionrate}), where we can see that $w_0$ appears only in the sum with $w_a$. 
Since the only geometrical constraint is on the sum $w_0+w_a<0$, non-standard cosmological models cannot be ruled out with these data alone. 
The highly bent shape of the contours can be explained further by the green line
in Fig.~\ref{BAO_models_comp}, which shows that the angular diameter distance for a cosmological model
with $\Omega_{\rm m}=0.5$ is departing significantly from the reference model at redshifts $z>0.5$. 

Although the total area of the confidence level contours is smaller
for the MCMC forecasts, the variance of the dark-energy parameters is
underestimated by the Fisher matrix also in this case. This introduces a bias on the DETF FoM of a factor of seven, as also shown in Table~\ref{tab:sigma_FM_MCMC}.

\begin{figure*}[hbtp]
\begin {center}
\includegraphics[width=0.45\textwidth, clip, trim=0mm 0mm 40mm 0mm]{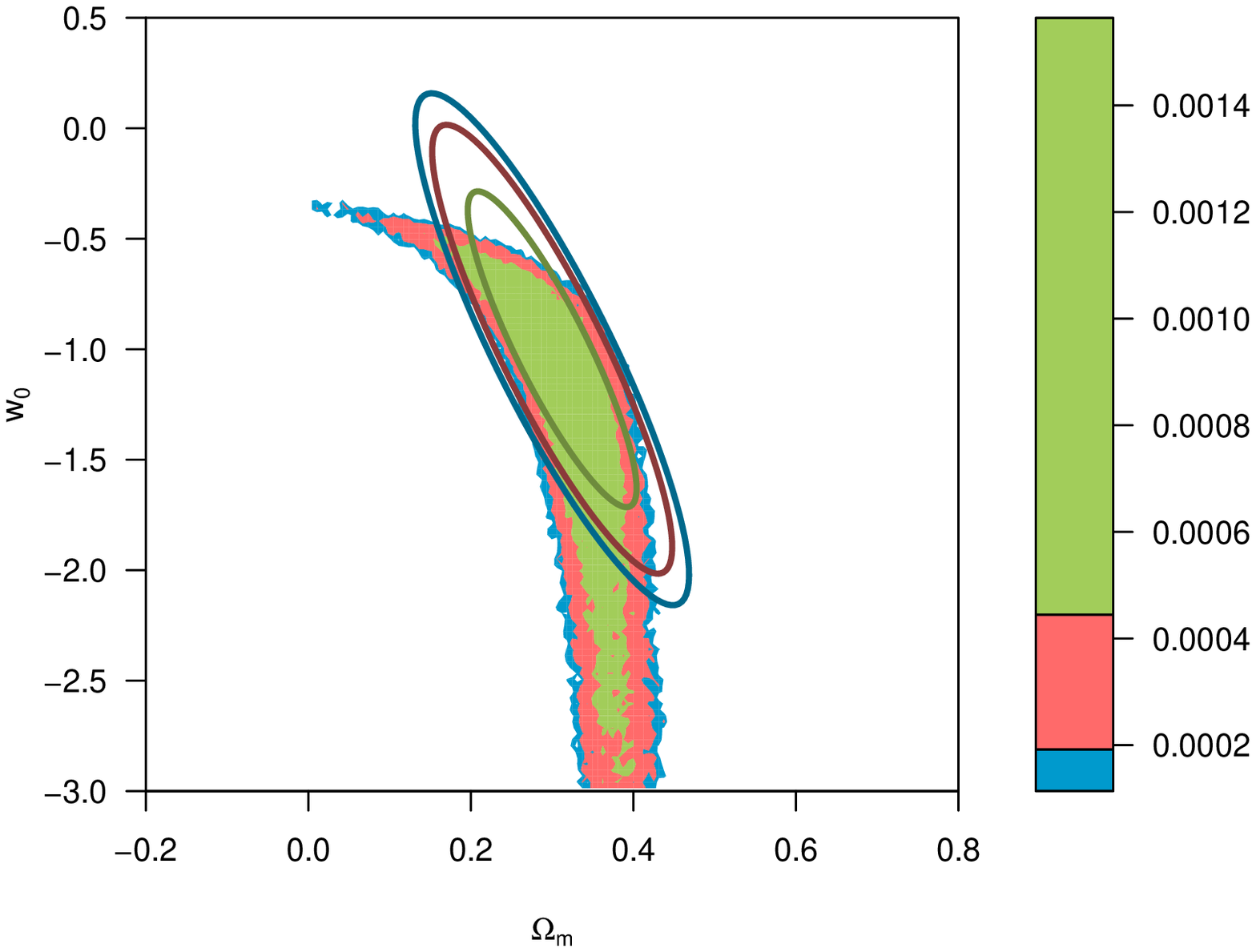}
 \includegraphics[width=0.45\textwidth,clip, trim=0mm 0mm 40mm
 0mm]{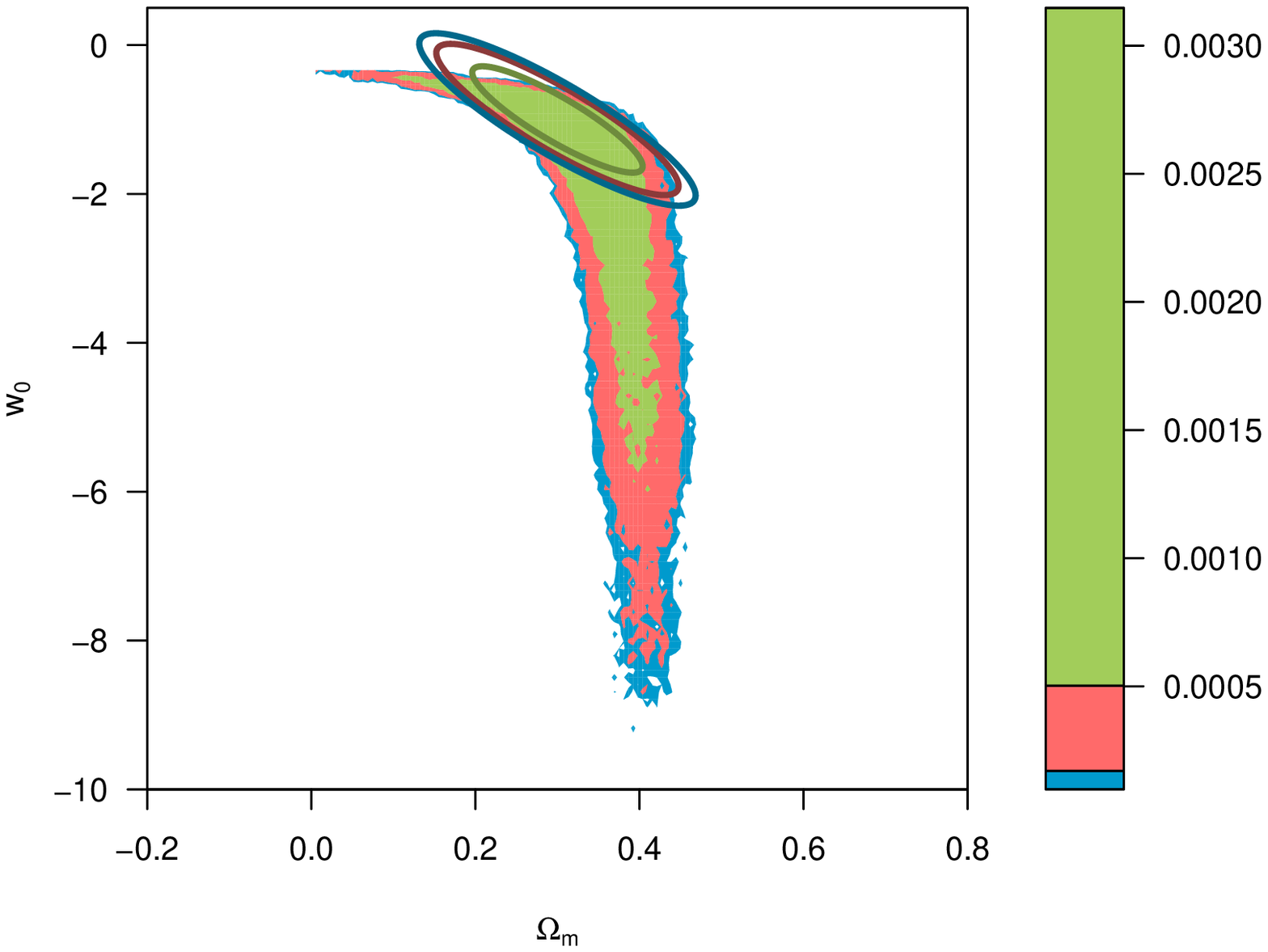}
 \caption{
  68\%, 90\% and 95\% confidence regions for a HETDEX-like BAO
  survey. The lower prior limit on $w_0$ is $-3$ (\emph{left panel})
  and $-10$ (\emph{right panel}), respectively. No further free
  parameters are used. Filled contours correspond to the MCMC, while the solid lines represent the Fisher matrix
   results.}

\label{baohetdex_omm_w0}
\end{center}
\end{figure*}

We show in Fig.~\ref{baohetdex_omm_w0} the forecasts for a HETDEX-like stage-III survey. As indicated in Table~\ref{tab:surveysettings}, as HETDEX is a
high-redshift survey with relatively small sky coverage, it will not be able to
put tight constraints on $w_a$, since in the CPL parameterisation dark
energy is not a dominant component of the Universe at
this epoch. In our forecasts we vary the parameters $\{\Omega_{\rm
  m}, w_0\}$ with flat priors $w_0>-3.0$ (left panel of Fig.~\ref{baohetdex_omm_w0}) 
and $w_0>-10$ (right panel), respectively. These plots give further evidence that the Fisher
matrix underestimates the variances of the dark-energy
parameters. Since $w_a=\text{const}=0$, the contours bend towards
negative values of $w_0$, to a much larger extend than for SNIa
(compare to the left panel of Fig.~\ref{BAO_SKA_w0_wa}).
The Fisher matrix is not able to predict the long tail of the posterior
distribution, even for relatively moderate priors on $w_0$.

\subsection{Weak lensing}
\label{sec:wl}

Weak gravitational lensing is the distortion of images of distant galaxies
caused by intervening matter along the line of sight between the source and the
observer. Weak lensing is sensitive to both the geometry of the Universe, and the
distribution of matter forming the large-scale structure. By measuring the
weak-lensing distortions from galaxies as a function of their redshift (lensing tomography),
information about the growth of structure can be deduced \cite{Bartelmann:2001p2082}.

We model two large tomographic weak-lensing surveys with DES-like and
Euclid-like settings, respectively.
The 2D projected convergence power spectrum $C_{\ell}^{\kappa_i \, \kappa_j}$ between redshift bins
$i$ and $j$ is given at any multipole $\ell$ by an integral over the three-dimensional matter power
spectrum $P_{m}(z,k)$ \cite{1992ApJ...388..272K, 1998ApJ...498...26K},
\begin{equation}
C_{\ell}^{\kappa_i \, \kappa_j} = \frac{9H_0^4\Omega_{\rm m}^2}{4c^3}
\int {\rm d} z\, \frac{g_{i}(z)g_j(z)
  (1+z)^2}{H(z)} \, P_m \left(z,\frac{\ell}{\chi(z)}\right) \, ,
\label{shearpowerspectrum}
\end{equation}
where the wavenumber can be set to $k = \ell / \chi(z)$, given the comoving distance
$\chi(z)$, by using the Limber approximation, which is accurate on all but the largest scales.
 The lens efficiency factor $g_i(z)$ is the equivalent of
the geometrical ratio of the lens--source to observer--source distances,
weighted with the galaxy distribution of the corresponding redshift
bin.
For simplicity, we ignore photometric redshift errors, and we
write the lens efficiency as an integral over the redshift bin width:
\begin{equation}
  g_i(z)=\int_{z_i}^{z_{i+1}} {\rm d}z' \, \left[ 1-\frac{\chi(z)}{\chi(z')}\right]
  \, \varphi(z') .
\end{equation}
Here $\varphi(z)$ represents the full galaxy redshift distribution, which is taken as \citep{Baugh:1997nc}
\begin{equation}
\varphi(z) \propto \left(\frac z {z_0}\right)^{\alpha}e^{-\left(\frac z
    {z_0}\right)^\beta}; \;\;\; \alpha = 2, \beta = 1.5,
z_0= z_{\rm med}/1.4 \, ,
\end{equation}
where the median redshift $z_{\rm med}$ is given in Table
\ref{tab:surveysettings} for both survey settings. The redshift
distribution is split into $N$ redshift bins between $z_{\rm min}$ and
$z_{\rm max}$, such that the number of galaxies is the same in each
bin (see Table~\ref{tab:surveysettings} for the values corresponding
to DES and Euclid).

We model the linear three-dimensional matter power spectrum with
spectral index $n_{\rm s} = 0.96$ and the
transfer function from Ref.~\cite{bbks86}. Baryonic damping is approximated
by modifying $\Gamma \equiv \Omega_{\rm m} h$ according to Ref.~\cite{1995ApJS..100..281S}.

We use the convergence power spectrum in the multipole range $0 < \ell < 1000$. On
the smallest of these scales, the density power spectrum is already in the mildly
non-linear regime. However, we choose the linear power spectrum for computational
ease, and since the comparison of forecasting methods will not be affected as long as we use the same assumption for both cases.
On intermediate scales, $ 100 < \ell < 1000$, where cosmic
variance is already non-Gaussian, and shot noise not yet dominating,
the non-Gaussian contribution to the covariance is not negligible. The errors increase and
mode-coupling between different multipoles is introduced \cite{2010A&A...514A..79P}.
However, the resulting underestimation of cosmological parameters is only at the
10\% to 20\% level \cite{2009MNRAS.395.2065T}.

\begin{figure*}
\begin {center}
  \includegraphics[width=0.9\textwidth]{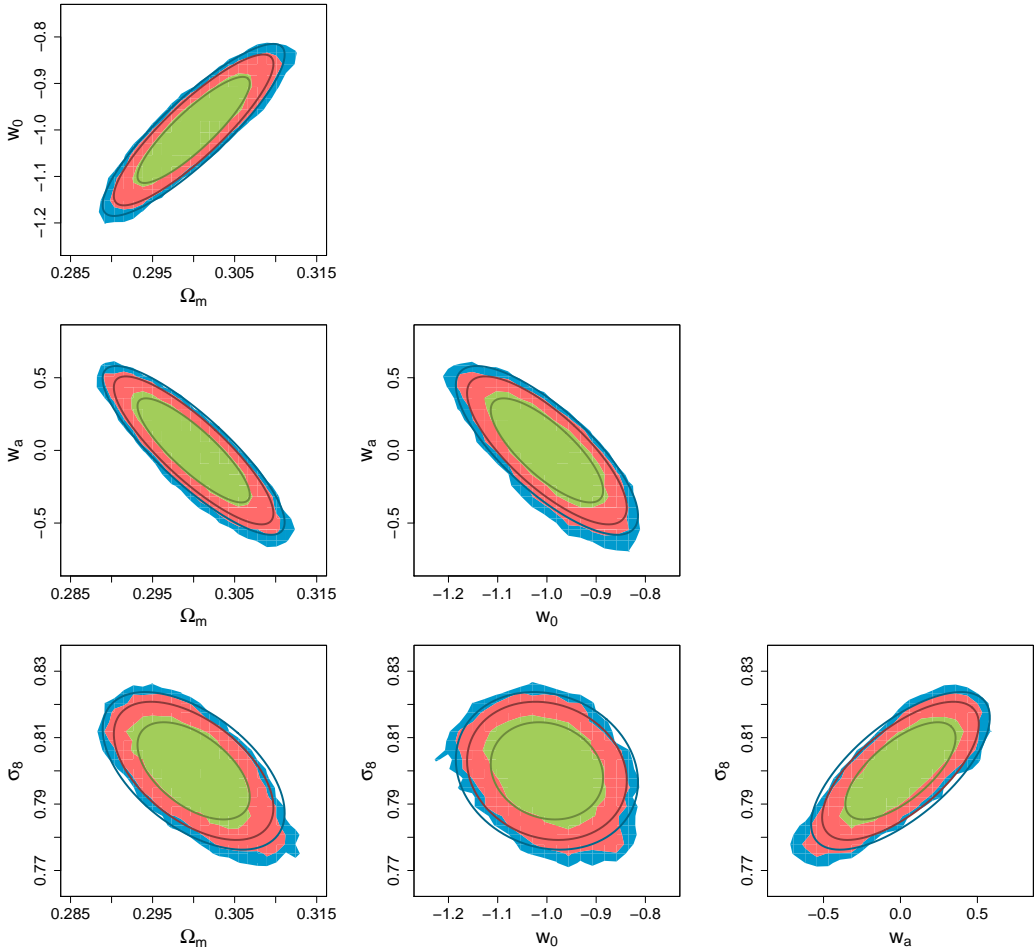}
  \caption{Marginalised 68\%, 90\% and 95\% confidence regions for a
    DES-like weak-lensing survey. Filled contours correspond to the MCMC, while the solid lines represent the Fisher matrix
   results.}
  \label{fig:WL_DES_CL}
\end{center}
\end{figure*}

\begin{figure*}
\begin {center}
  \includegraphics[width=0.9\textwidth]{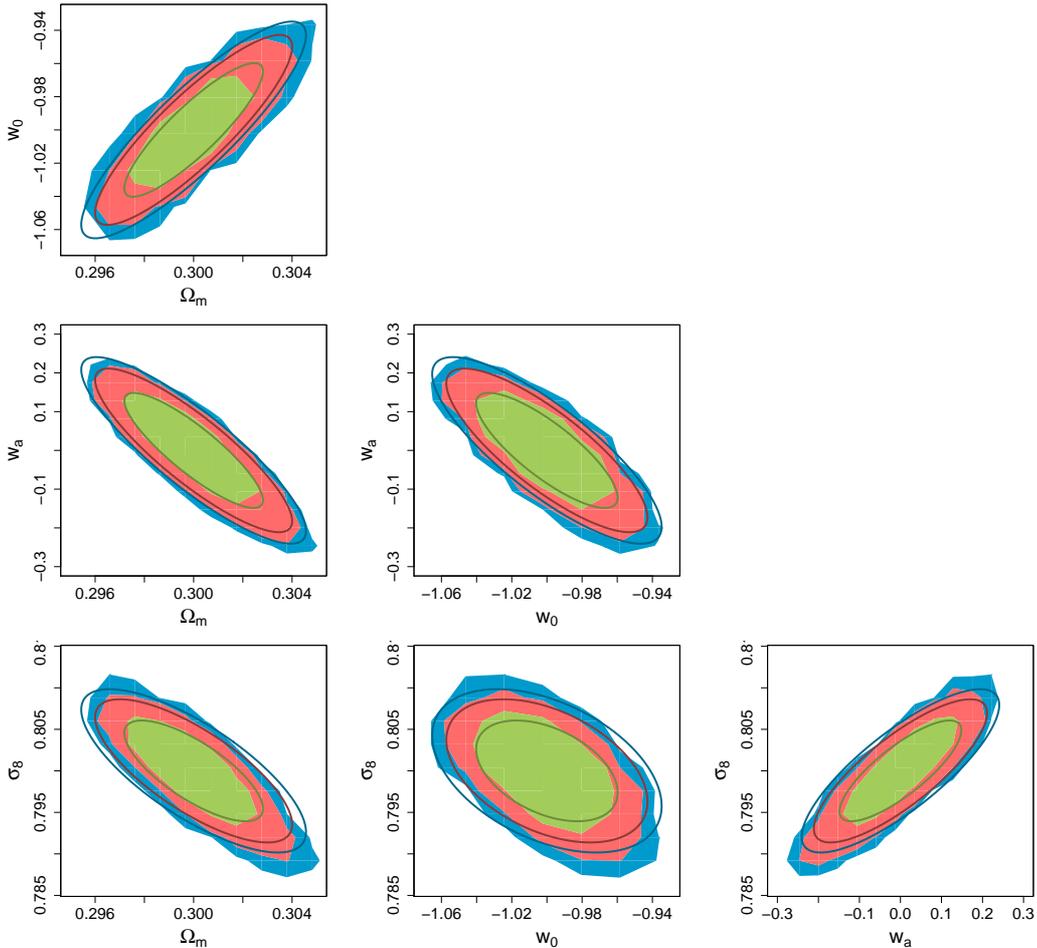}
  \caption{Marginalised 68\%, 90\% and 95\% confidence regions for a
    Euclid-like weak-lensing survey.}
  \label{fig:WL_Euclid_CL}
\end{center}
\end{figure*}

The comparison of the Fisher ellipses with full likelihood contours
for a DES- and Euclid-like survey are shown in
Figs.~\ref{fig:WL_DES_CL} and \ref{fig:WL_Euclid_CL},
respectively. The results of the method comparison for both surveys are similar, and the
shapes and sizes of the confidence contours are in this case in agreement for the two methods, since the
likelihood function features fairly elliptical contours for all
parameter combinations. These results are in accordance with a previous analysis in \cite{Taylor:2010pi}.

\section{Towards Gaussian Parameters}
\label{sec:gauss}

We have shown in the previous sections that the Fisher matrix technique can have severe
weaknesses in predicting sizes and shapes of confidence level contours. The contours predicted by distance
measure probes are not well-described by the Fisher matrix. The degeneracy of the
dark-energy parameters $w_0$ and $w_a$ in the CPL parameterisation in the expansion
rate cannot be broken, and therefore a Gaussian approximation of the likelihood is not
valid. For this reason, we introduce a new parameterisation, and suggest that SN and
BAO Fisher matrix forecasts based on distance measures only should be considered in this
new parameter space.

\begin{figure*}
\begin {center}
  \includegraphics[width=0.45\textwidth,clip, trim=0mm 0mm 40mm 0mm]{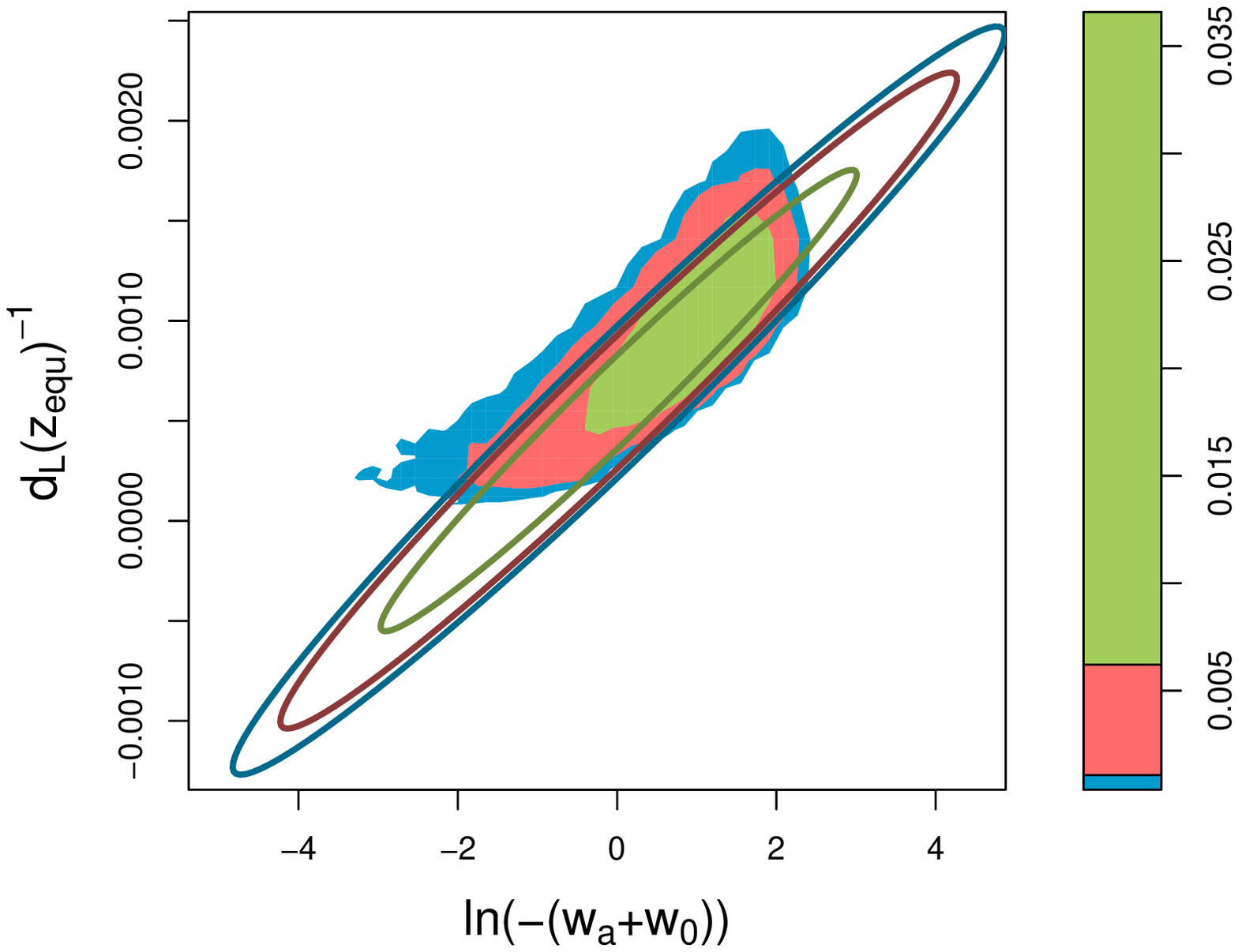}
  \includegraphics[width=0.45\textwidth,clip, trim=0mm 0mm 40mm 0mm]{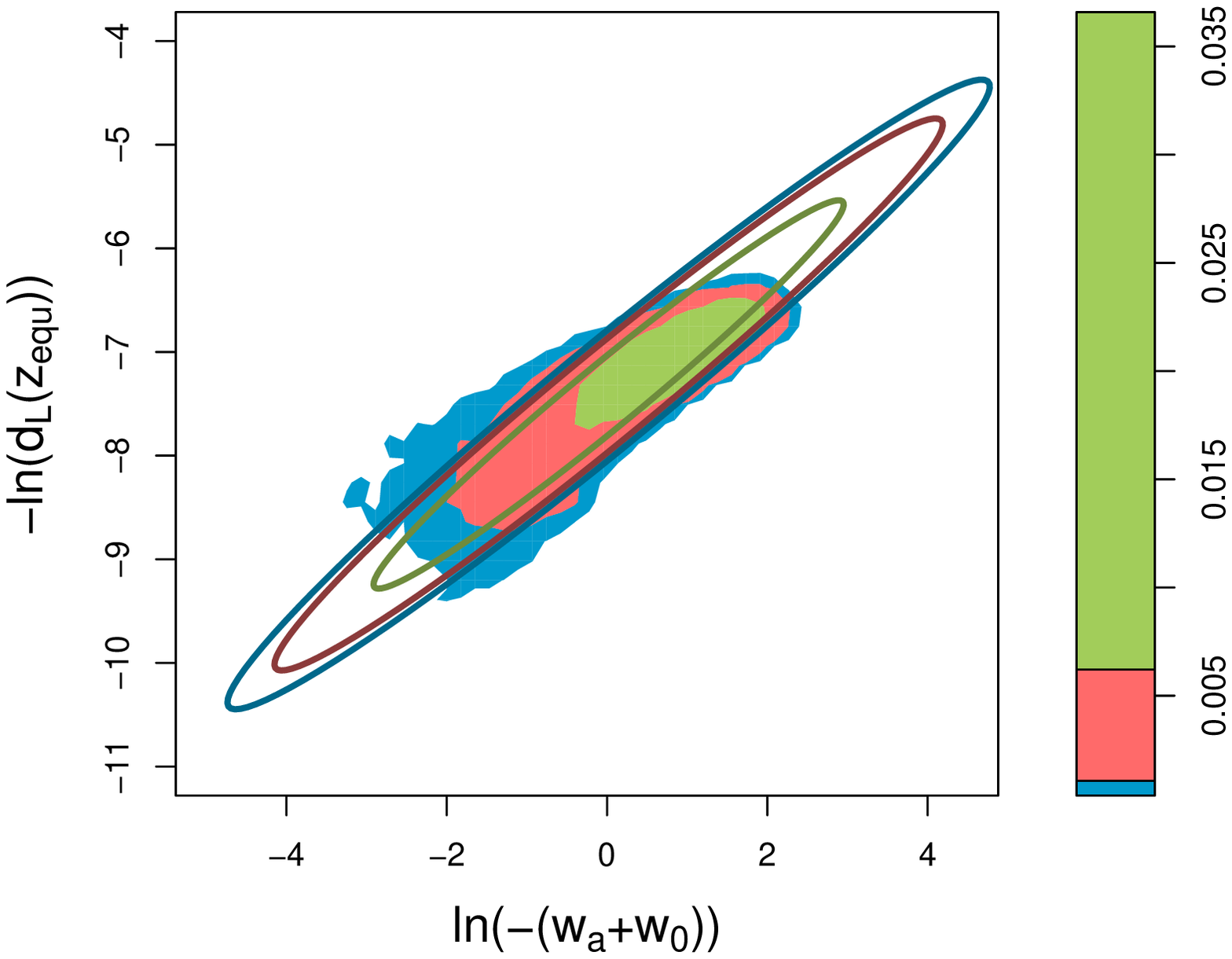}

  \caption{68\%, 90\% and 95\% confidence regions in new parameterisations, marginalised over $\Omega_{\rm m}$, for a supernova survey. Filled contours correspond to the MCMC, while the solid lines represent the Fisher matrix
   results.}
  \label{fig:SN_newpara}
\end{center}
\end{figure*}

First, we define a new parameter $w_s \equiv \ln [-(w_0+w_a)] $ in order to exclude the region
with $w_0+w_a>0$, 
which is clearly ruled out by the MCMC, as shown in
Figs.~\ref{SN_omm_w0_notmarg} and \ref{BAO_SKA_w0_wa}. Motivated by
the primary parameterisations of the BAOs and CMB, we introduce the luminosity
distance $d_{\rm L}(z_{\rm equ})$ as the basis of a new
parameter. However, unlike in the case of CMB, there is
no unique redshift for SN, at which the observables
are measured. The only scale which is singled out in the redshift
  domain is the redshift of dark-matter and dark-energy equality $z_{\rm equ}$, which is defined by
\begin{equation}
\Omega_{\rm m} (1+z_{\rm equ})^{3} = \Omega_{\Lambda} e^{-3w_a\frac z{1+z_{\rm
      equ}}}(1+z_{\rm equ})^{3(1+w_0+w_a) }.
\end{equation}
We transform the MCM chain into the new parameter space $\{ w_s,
{d_{\rm L}}(z_{\mathrm{equ}}) \}$ and perform a new Fisher
matrix calculation. We restrict our analysis to the SN probe since SN and BAO show
the same characteristics, and are based on the same analytical expressions.  The bare
luminosity distance at $z_{\rm equ}$ is still near-degenerate with
$w_s$; therefore we apply another non-linear transformation by taking the
inverse and the logarithm, respectively. Taking the logarithm ensures the positive definiteness of the luminosity distance. The result is shown in
Fig.~\ref{fig:SN_newpara} for the new parameters $\boldsymbol \theta_1 = \left\{ w_s , \,
  d_{\rm L}^{-1}(z_{\rm equ})\right\}$ and
$\boldsymbol \theta_2 = \left\{ w_s, \, -\ln \left[ d_{\rm L}(z_{\rm equ}) \right] \right\}$,
respectively.

The luminosity distance $d_{\rm L}(z_{\rm equ})$ depends via $z_{\rm equ}$ on
$\Omega_{\rm m}$, $w_0$ and $w_a$. However in the process of marginalisation over one
parameter, in this case $\Omega_{\rm m}$, information is lost. This implies the
irreversibility of the transformation and we have to compare models
with the likelihood contours in this transformed parameter space.

\section{Discussion and Conclusions\label{sec:conclusion}}

\begin{table}
\begin {center}
\begin{tabular}{|l|c|c|c|c|}
 \hline
 Survey & $\delta_{\Omega_{\rm m}}$ & $\delta_{w_0}$ & $\delta_{w_a}$ &   $\text{FoM}_\text{Fisher}/\text{FoM}_\text{MCMC}$\\
 \hline
 SN         & 1.43 & 0.40 & 0.86  & 1.20\\
 BAO SKA    & 2.52 & 0.308 & 0.463 & 0.74\\
 BAO HETDEX & 0.62 & 0.38 & --- & --- \\
 WL DES     & 0.934 &0.940 & 0.921 & 1.30\\
 WL Euclid  & 1.01 & 1.04 & 0.968  & 1.12\\ \hline
 \end{tabular}
 \caption{For each of the three parameters $\Omega_{\rm m}, w_0$ and $w_a$, and for the five
   surveys settings,
   the values show the ratio of marginalised variances for the Fisher matrix over the MCMC, $\delta_{\Omega_{\rm m}}$, $\delta_{w_0}$ and $\delta_{w_a}$ respectively.
   A ratio smaller than unity means that the Fisher matrix underestimates the variance of the full
   posterior distribution. The fourth column shows the ratio of the DETF FoM, where we defined the FoM as the inverse of the area of the 1-$\sigma$ contour.}
 \label{tab:sigma_FM_MCMC}
\end{center}
\end{table}

We have investigated the comparison of the two forecast methods under the following four
aspects: their practicality in application, the morphology of the marginalised
confidence contours, the impact on the DETF FoM, and the Gaussianity of
parameter combinations.

\begin{itemize}
\item \emph{Practicality:} The advantage of the Fisher matrix forecast over
  MCMC is the computational speed. However, the Fisher matrix relies
  on the stability of numerical derivatives of the observables or the
  likelihood. To ensure high enough accuracy, it is necessary to probe
  many points corresponding to different step sizes $h$ for the
  numerical derivative (adaptive or non-adaptive). Furthermore,
  compared to the requirements for MCMC, a much higher numerical
  accuracy for each of the cosmological quantities is required. This
  is true in particular for parameters which have a subtle effect on
  the likelihood, such as $w_a$. As a consequence, the tabulation of
  recurrent quantities (such as the growth factor or the angular
  diameter distance) as well as the numerical integrations have to be
  undertaken with much higher precision. This will result in a
  increased computation time for one model by a factor of many. Fisher
  matrix forecasts will be even more demanding for more complex and
  more realistic calculations, for example, using non-linear models of
  the large-scale structure, full Boltzmann integrator for the
  transfer function, departures from the Limber approximation of the
  power spectrum, and correlations between cosmological probes
  \cite{2011arXiv1109.0958G}.

\item \emph{Morphology:} On one hand, for purely geometrical probes such as supernovae, and more evidently BAOs, the
  likelihood function is not elliptical, and is asymmetric. Extended degeneracies exist between the parameters, since any combination thereof which reproduces the correct distance measure is allowed. However, other regions are excluded:
for example firstly, if the sum
  $w_0 + w_a$ is positive, then dark energy at late times scales faster
  than matter, which is disfavoured by the data. Secondly, the likelihood is
  zero for a negative matter density parameter $\Omega_{\rm m}$. The
  likelihood of the Fisher Gaussian approximation reaches well into these
  `forbidden' regions, while being naturally unable to trace the extended degeneracies which in reality exist.
  The existence of such forbidden regions prevents us from defining parameter transformations which
  produce fully Gaussian likelihood distributions. An asymmetry due to
  physically forbidden regions always remains, making the Fisher
  matrix not a good approximation of the distribution.

On the other hand, for more complex observables which also
trace structure formation, such as in our case weak lensing, all
cosmological parameters enter the likelihood calculation in multiple
ways: not only through the Hubble expansion, but also via the
transfer and growth functions. This means that degeneracies are
significantly alleviated, as we have shown in
Figs.~\ref{fig:WL_DES_CL} and \ref{fig:WL_Euclid_CL}. The same is likely to happen for a full tomographic galaxy clustering analysis, and especially so when additional features, such as redshift-space distortions and the Alcock-Paczynski effect, are included in the calculation \cite{2003ApJ...598..720S,2011arXiv1109.0958G}.

\item \emph{Impact on the figure of merit:} The ratio of the variances
  estimated from the Fisher matrix to the one corresponding to MCMC, as shown in Table~\ref{tab:sigma_FM_MCMC}, is expected to be smaller than unity in
  order to satisfy Eq.~(\ref{cramer_rao}). With the exception of the ratios
  for $\Omega_{\rm m}$ for SN and BAO SKA (as discussed in
  Section~\ref{sec:BAO}), this is indeed the case. The
  largest mismatch using the Fisher matrix occurs for the purely
  geometrical probes, and in particular for the SKA
  BAOs. Interestingly the deviations in the Figure of Merit, which we 
  define as the inverse of the area of the 68$\%$ likelihood contour, is not as large
  as the deviations on the single errorbars and morphologies of the contours. This is due to the
  different shape of the Gaussian contours, which often reach in an
  area, where the true likelihood is vanishing, and then
  underestimating the errors in other ranges. In addition, we should
  note that, the estimation of the area of the MCMC likelihoods has
  been performed on a relatively coarse grid, so it is also
  overestimated, this is in particular the case for the high accuracy WL surveys. 
  Hence, with the exception of BAO SKA, we find the FoM
  in agreement on the 20$\%$ level with the Fisher matrix forecasts.
  
\item \emph{Gaussianity:} There have been previous studies on the Gaussianity of parameter
  combinations. For example, Ref.~\cite{Joachimi:2011iq} lifted the degeneracy
  of parameters using the Box-Cox transformation \cite{box1964analysis}: they linearised
  the parameters in the $\Omega_{\rm m}$--$\sigma_8$ plane for a future
  weak-lensing survey. The approach worked well for the slightly
  curved contours in their case but, in the case of highly
  curved, non-Gaussian likelihood shapes such as for SN
  (see Fig.~\ref{SN_omm_w0_notmarg}), we find that Box-Cox
  transformations do not provide a sufficient linearisation of the
  parameter to produce a Gaussian-like distribution.

  Further, in Ref.~\cite{Hawken:2011nd} a comparison of
  forecasting methods to probe the large-scale structure is performed. They also
  found major differences of Fisher matrix and MCMC for small-volume
  surveys.

  In this work, we have explored new, more Gaussian parameterisations,
  which are physically motivated. In this way, we succeed in linearising the parameter
  degeneracies, but the resulting likelihood function is still sufficiently
  non-Gaussian, that the Fisher matrix approximation remains
  relatively poor.
  In addition it is not possible to use the
  transformation we suggest to make the joint two-dimensional
   constraints on $w_0$ and $w_a$ more representative with the
   Fisher matrix approach. The reason for this is that after
   marginalisation over the other parameters, it is not possible to
   take the inverse of the transformation, since it also depends on
   $\Omega_{\rm m}$. Of course it would be possible to perform a
   numerical marginalisation (integration) in the full 3-dimensional
   back-transformed parameter constraints. But then the question
   arises how advantageous the Fisher matrix approach is in the
   first place, and whether one is not better off by performing a MCMC
   analysis. However, there might still a possibility that there is a
   useful parameterisation just involving $w_0$ and $w_a$.

\end{itemize}

We
conclude from our comparisons that the Monte Carlo sampling analysis is preferable in the purely geometric cases, for which the Fisher
matrix method has significant weaknesses in predicting parameter
constraints regarding the morphology and orientation of the degeneracies. 

When using simple probes of the expansion history, such as supernovae, the long parameter degeneracies are missed by the Fisher approach, while the problem is alleviated by the use of more complex tests of structure formation, such as weak lensing, for which we confirm previous results on the Gaussianity of the posterior probability.
We highlight that we have not explicitly included CMB priors
on the parameters, nor combined the probes with the WMAP (or Planck) likelihood. Although WMAP or
Planck parameter constraints have no significance on the dark-energy
parameters $w_0$ and $w_a$, they are able to break degeneracies
with respect to $\Omega_{\rm m}$. Naturally, the combination of probes will
tighten all constraints and make the likelihoods more Gaussian.
However, in order to explore the merits of a single probe, we
deem such a procedure misleading. 
Moreover, in general also the CMB likelihood deviates from
Gaussianity and strong parameter degeneracies can be present \cite{Perotto:2006}.

In conclusion we 
advocate the full MCMC likelihood forecast method in order to explore realistically
the merits of future dark energy surveys with geometrical probes of the expansion history; more generally the full MCMC method should be applied for any setting (combination of model parameters and observables), in which the Gaussianity of the likelihood distribution is unknown.

\acknowledgments We would like to thank Sarah Bridle, Henk Hoekstra, Tom Kitching, Michael Kopp and Eric Linder for useful discussions. The work of MK, JW and TG is supported by the DFG TRR 33 `The Dark Universe'.

\bibliographystyle{JHEP}
\def\apj{Ap.\ J.}  
\def\mnras{MNRAS} \def\mn{MNRAS}
\def\apjl{Ap.\ J.~Letters}  
\def\apjs{Ap.\ J.~Suppl.}  
\def\prd{Phys.~Rev. {\bf D}}
\def\aap{A\&A}
\def\jcap{JCAP}
\newcommand{\prl}{{Phys.~Rev.~Lett.}}
\providecommand{\href}[2]{#2}\begingroup\raggedright\endgroup

\end{document}